# Temperature Dependent Low-Frequency Noise Characteristics of NiO$_x$/β-Ga$_2$O$_3$ *p-n* Heterojunction Diodes

Subhajit Ghosh[1], Dinusha Herath Mudiyanselage[2], Fariborz Kargar[1], Yuji Zhao[3], Houqiang Fu[2], and Alexander A. Balandin[1,4,×]

[1]Department of Electrical and Computer Engineering, University of California, Riverside, California 92521 USA

[2]School of Electrical, Computer, and Energy Engineering, Arizona State University, Tempe, Arizona 85287 USA

[3]Department of Electrical and Computer Engineering, Rice University, Houston, Texas 77005, USA

[4]Department of Materials Science and Engineering, University of California, Los Angeles California 90095 USA

[×] Corresponding author (AAB) e-mail: balandin@seas.ucla.edu

**Abstract**

We report on the temperature dependence of the low-frequency electronic noise in $NiO_x/\beta$-$Ga_2O_3$ *p-n* heterojunction diodes. The noise spectral density is of the *1/f*-type near room temperature but shows signatures of Lorentzian components at elevated temperatures and at higher current levels (*f* is the frequency). We observed an intriguing non-monotonic dependence of the noise on temperature near $T = 380$ K. The Raman spectroscopy of the device structure suggests material changes, which results in reduced noise above this temperature. The normalized noise spectral density in such diodes was determined to be on the order of $10^{-14}$ cm$^2$/Hz (*f* = 10 Hz) at 0.1 A/cm$^2$ current density. In terms of the noise level, $NiO_x/\beta$-$Ga_2O_3$ *p-n* diodes occupy an intermediate position among devices of various designs implemented with different ultra-wide-bandgap semiconductors. The obtained results are important for understanding the electronic properties of $NiO_x/\beta$-$Ga_2O_3$ heterojunctions and contribute to the development of noise spectroscopy as the quality assessment tool for new electronic materials and device technologies.

**Keywords:** *1/f* noise; flicker noise; wide bandgap semiconductors, *p-n* diodes

## 1. Introduction

In recent years there has been a growing interest in innovative semiconductor heterostructures and electronic technologies to address the ever-increasing industry demands[1–6]. Ultra-wide-bandgap (UWBG) semiconductor materials have attracted interest for device applications in high-power electronics[7–9]. Materials such as AlGaN, AlN, $Ga_2O_3$, BN, and diamond emerged as viable options for future power electronic materials and well-established wide-bandgap technologies such as GaN and SiC. In particular, β-$Ga_2O_3$ has motivated significant research interest owing to easily available high-quality β-$Ga_2O_3$ substrates and the materials' promising electrical and optical characteristics[10,11]. Essential device types, including field effect transistors (FETs), and Schottky barrier diodes (SBDs) have been demonstrated[12,13]. Obtaining *p*-type $Ga_2O_3$ to form *p-n* junction bipolar diodes is difficult[14]. For this reason, $NiO_x$ has been used as a *p*-type material to create $NiO_x$/β-$Ga_2O_3$ *p-n* heterostructures[15]. Bulk $NiO_x$ has a cubic (NaCl-type) structure with a lattice parameter of 0.4177 nm[16] and a bandgap, ranging from 3.6 eV to 4.0 eV[17]. It is known that $NiO_x$ is a natural *p*-type UWBG semiconductor, which can be easily deposited into other *n*-type semiconductor materials[18,19]. Such $NiO_x$-based heterostructures have been used in *p-n* diodes and other devices[20,21]. Previous studies on $NiO_x$/β-$Ga_2O_3$ heterostructure diodes reported their promising characteristics, including low leakage currents, high on-off ratio, and low on-resistance[22–24]. Interestingly, $NiO_x$ is also a well-known spintronic material with a high Néel temperature of 523 K for bulk crystals[25].

In this Letter, we report on the low-frequency noise (LFN) characteristics of $NiO_x$/β-$Ga_2O_3$ *p-n* heterojunction diodes, focusing on their temperature dependence. The noise measurements are important for understanding the electronic characteristics of $NiO_x$/β-$Ga_2O_3$ heterojunctions, particularly at elevated temperatures. Since UWBG devices are designed to operate at elevated temperatures, the knowledge of high-temperature noise spectra gains extra significance. The most common noise types in semiconductor materials include *1/f*-type flicker noise and Lorentzian-shaped generation-recombination (G-R) noise[26,27]. The noise data provides information on charge carrier dynamics and defects, acting as trapping centers. In the case of semiconductor heterostructures, low-frequency noise sheds light on the effect of the interfacial states at heterojunctions on the device's performance. In $NiO_x$/β-$Ga_2O_3$ heterostructures, the interfacial states emerge due to the dangling bond density of the β-$Ga_2O_3$ substrate plane, which impacts the device performance. Our results contribute to the development of noise spectroscopy as the quality assessment tool for new UWBG device technologies.

## 2. Experimental Section

For this study, we utilized NiO$_x$/β-Ga$_2$O$_3$ *p-n* heterostructure diodes on ($\bar{2}$01) β-Ga$_2$O$_3$ substrate. This type of structure was selected because NiO$_x$/β-Ga$_2$O$_3$ heterostructures fabricated on ($\bar{2}$01) substrates have shown high-quality Ohmic contacts, lower turn-on voltage, and better ideality factors. To fabricate the devices, first, ($\bar{2}$01) β-Ga$_2$O$_3$ substrates were obtained from Novel Crystal Technology, Inc, Japan. To clean the substrate, a standard cleaning procedure was implemented, which included sequential cleaning with acetone, isopropyl alcohol, and deionized water, aided by sonication. Next, the back contacts of Ti/Au (20/130 nm) were deposited using electron beam (E-beam) evaporation, followed by rapid thermal annealing at 500 °C in an N$_2$ environment. Subsequently, standard photolithography techniques were employed to define circular patterns for the deposition of NiO$_x$ and the anode. Using E-beam evaporation, layers of 200 nm NiO$_x$ and the anode Ni/Au (20/130 nm) were deposited, followed by a lift-off process. After that, the devices were subjected to 350 °C annealing in N$_2$ environment for 1 minute. This step was performed to improve the device performance by forming a high-quality Ohmic contact between the Ni/NiO$_x$ interface and reducing the number of interface states at the NiO$_x$/β-Ga$_2$O$_3$ heterojunction[28].

The temperature-dependent current–voltage (I–V) measurements were conducted in the vacuum inside a probe station (Lakeshore TTPX) in a 2-terminal configuration using a semiconductor parameter analyzer (Agilent B1500). During the measurements, the sample was placed on top of a sample stage and heated up to 395 K using a temperature controller (Lakeshore Model 336). Figure 1 (a) shows the forward bias I–V characteristics in the semi-log scale of a NiO$_x$/β-Ga$_2$O$_3$ *p-n* heterojunction diode with a 300 μm diameter, measured at RT ($T$ = 296 K). The ideality factor, *n* of the diode device was calculated to be ~1.9 in the lower current region. The temperature-dependent characteristics of the diodes were conducted at temperatures in the range from 300 K to 395 K, in the heating cycle. Figure 1 (b) shows the forward-bias I–Vs of the diode at different temperatures. The turn-on voltage, $V_T$, of the device, decreases linearly as the diode temperature increases. The decrease in $V_T$ at elevated temperatures is attributed to the decrease in the depletion width at the heterojunction owing to the thermal diffusion of holes from the *p*-NiO$_x$ layer[29].

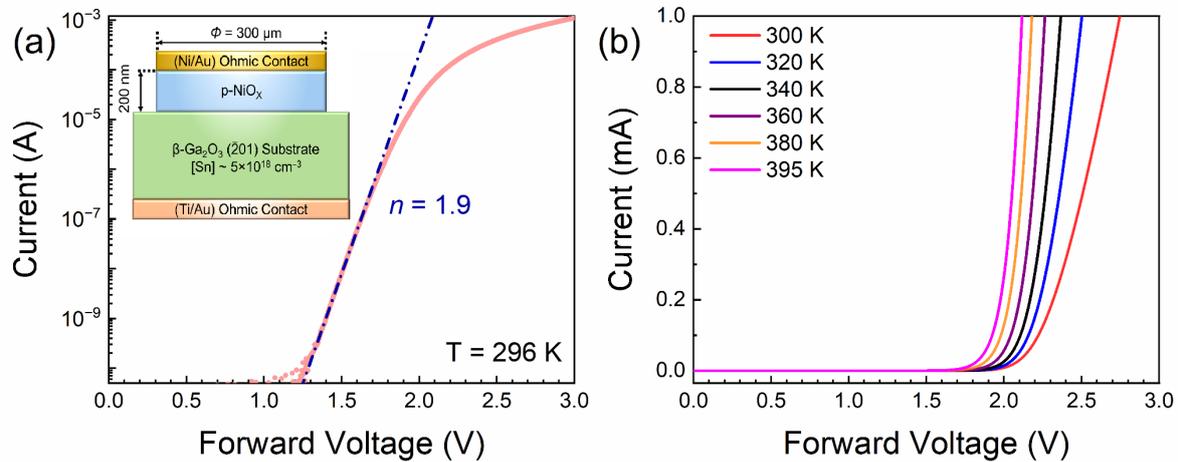

**Figure 1**: (a) Room temperature I–V characteristics of NiO$_x$/β-Ga$_2$O$_3$ *p-n* heterojunction diode in the semi-log scale. The extracted ideality factor of the diode is *n* = 1.9. The inset shows a schematic of the device. (b) The temperature-dependent I–V characteristics of the device measured between 300 K and 395 K. The turn-on voltage ($V_T$) shifts towards lower forward voltages with increasing temperature.

The low-frequency noise measurements were conducted inside the probe station chamber under a high vacuum following a standard protocol[30,31]. The noise measurement system consists of the device under test connected in series to a load resistor and powered by a low-noise DC battery used as a voltage source. A potentiometer was connected to control the voltage drop across the circuit. During the noise measurements at each temperature point, the output voltage fluctuation, *ΔV*, was transferred to the low-noise voltage preamplifier (SR 560) which amplified the signal. The amplifier was connected to a dynamic signal analyzer (Photon+) to convert the time domain voltage fluctuation signal to the corresponding frequency-dependent voltage spectral density, $S_V$. For noise data analysis, the obtained voltage spectral density, $S_V$, was converted to the corresponding current spectral density, $S_I$. Further details of our noise measurement protocol, in the context of other materials and devices, can be found in prior reports[30–32].

## 3. Results and Discussions

The LFN characteristics of the diode are presented in Figure 2 (a)–(d) at varying forward currents and different temperatures. Figure 2 (a) shows the current spectral density, $S_I$, as a function of frequency, $f$, at different diode current regimes at 300 K. The noise spectra show a consistent *1/f*, flicker-type noise dependence at all the measured currents. Similar *1/f* power law-dependent noise behavior was observed at varying device current levels at the next measured temperatures of 320 K and 340 K (refer to the supplemental Figure S2). The nature of the noise behavior starts to change at 360 K as shown in Figure 2 (b). In this temperature, the noise spectra at the lower current regimes followed *1/f* dependence as observed previously. However, the noise behavior changes to the *1/f²*-type noise at higher current levels (starting from $I = 5 \times 10^{-4}$ A). The *1/f²*-type spectrum indicates the tail of the Lorentzian, characteristic of the G-R noise. The Lorentzian can indicate either a dominant trap with a specific time constant or a phase transition[26]. To better understand the change in the noise characteristics with temperature and current we plotted the normalized current spectral density, $S_I/I^2$, as a function of frequency, $f$, at constant diode currents with varying temperatures. Figure 2 (c) presents $S_I/I^2$ for elevated temperatures at the fixed current level of $I = 1 \times 10^{-6}$ A. The noise shows *1/f* dependence at temperatures up to 360 K. At 380 K the noise spectrum changes to the *1/f²* dependence with the noise level increased by several orders of magnitude as compared to the noise at lower temperatures. At the next temperature of 395 K, the noise spectrum becomes the *1/f*-type again and decreases in the overall level. The non-monotonic dependence of noise on temperature was also observed at other current levels as shown in Figure 2 (d) for $I = 1 \times 10^{-5}$ A. A possible origin of this intriguing non-monotonic noise behavior with temperature is discussed below.

It is known that flicker noise is seldom exactly *1/f* type. More often its spectrum is described as $S_I \sim 1/f^\gamma$, where $\gamma$ is the noise parameter. The deviation of the $\gamma$ parameter from unity and its dependence on temperature and current are considered to be important metrics for semiconductor devices[26]. Figure 3 (a) provides the extracted $\gamma$ values as a function of the forward bias current, $I$, at different temperatures for a NiO$_x$/β-Ga$_2$O$_3$ *p-n* heterojunction diode. The $\gamma$ value varies between 0.9 – 1.3 across the measured currents at temperatures between 300 K and 340 K. At 360 K the $\gamma$ parameter sharply increases to ~2 at higher currents. At 380 K, the $\gamma$ parameter is between 1.9 and 2.3 owing to the emergence of the G-R features. The extracted $\gamma$ value drops to ~1.6 at 395 K which indicates a non-monotonic change in noise spectral shape. One can see in Figure 3 (b) that the $\gamma$ remains close to ~1 at lower temperatures up to 360 K, and low current levels. It increases sharply to ~2 at 380 K and drops again at 395 K. Previously, a strong dependence of the $\gamma$ parameter on bias voltage was interpreted as evidence of the non-uniform distribution of traps in space and energy[33,34]. We have a weak current dependence on the $\gamma$ parameter. The deviation $\gamma$ from unity and sharp increase at 380 K are likely indicating some changes in the material or transport regime as discussed below.

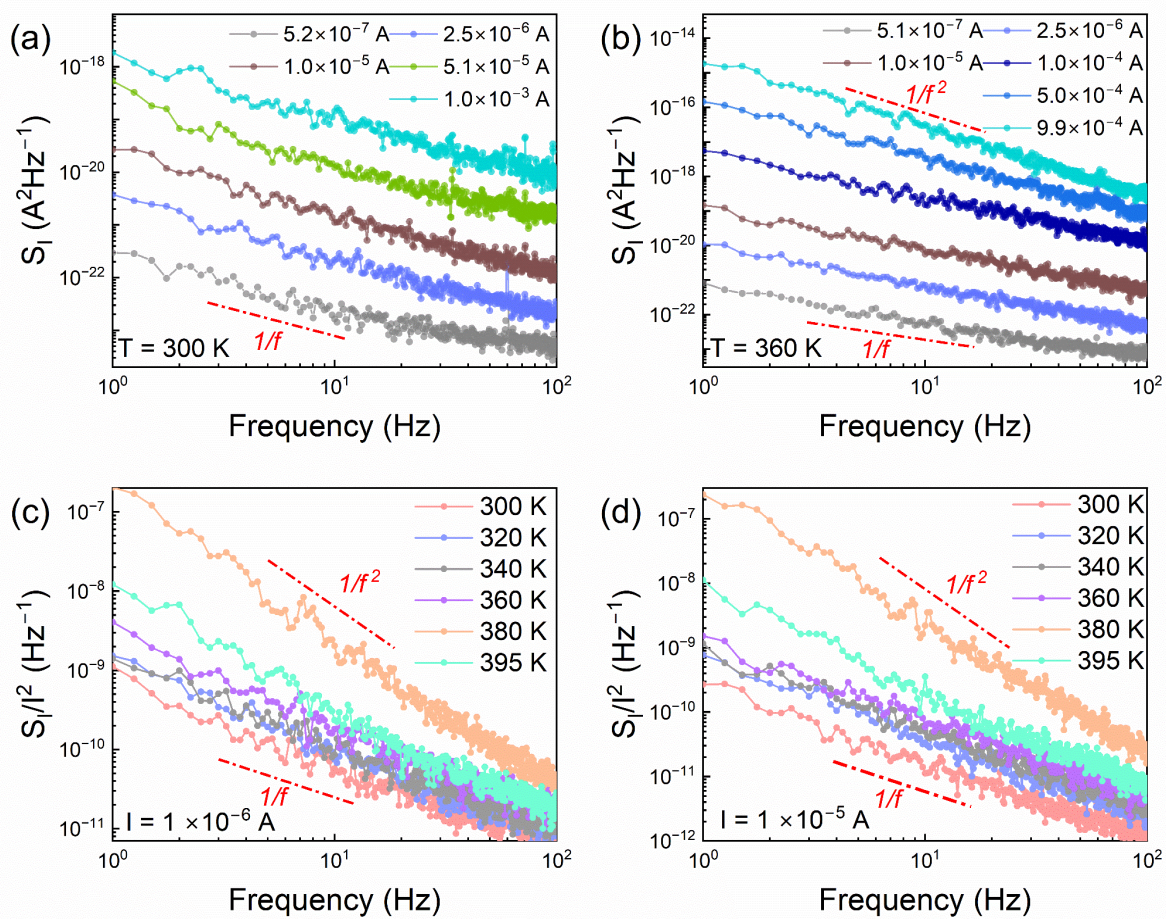

**Figure 2**: (a) The current noise spectral density, $S_I$, as a function of frequency, $f$, measured at $T$ = 300 K for different forward current values. The noise spectra are of $1/f$ type for all currents. (b) The $S_I$ vs $f$ characteristics for different currents at the elevated temperature of 360 K. The noise spectra remain $1/f$ at lower and intermediate current levels but change to $1/f^2$ Lorentzian type at higher currents, above $I$ = 100 μA. (c) The normalized noise spectral density, $S_I/I^2$, as a function of frequency, $f$, at current $I$ = 1 μA at different temperatures. The noise spectra change from $1/f$ at 300 K to $1/f^2$ at higher temperatures. (d) The corresponding temperature-dependent $S_I/I^2$ vs $f$ for the current $I$ = 10 μA.

In Figure 4 (a) we present the current spectral density, $S_I$, measured at $f = 10$ Hz as a function of forward current, $I$, at different temperatures. At smaller temperatures, the noise behavior follows $S_I \sim I$ linear dependence. Such noise behavior is typical for diode devices and has previously been observed for other diode technologies[30,35,36]. Interestingly, $S_I(I)$ dependence becomes closer to quadratic, $S_I \sim I^2$, at 380 K. The quadratic behavior is expected for linear resistors and can also be found in diodes in certain transport regimes[31,32,37,38]. Figure 4 (b) shows the normalized noise spectral density, $S_I/I^2$, at $f = 10$ Hz as a function of temperature for different currents. At lower currents, the normalized noise level remains within a specific range till 360 K and increases sharply to its maximum value at 380 K before dropping to lower levels at 395 K.

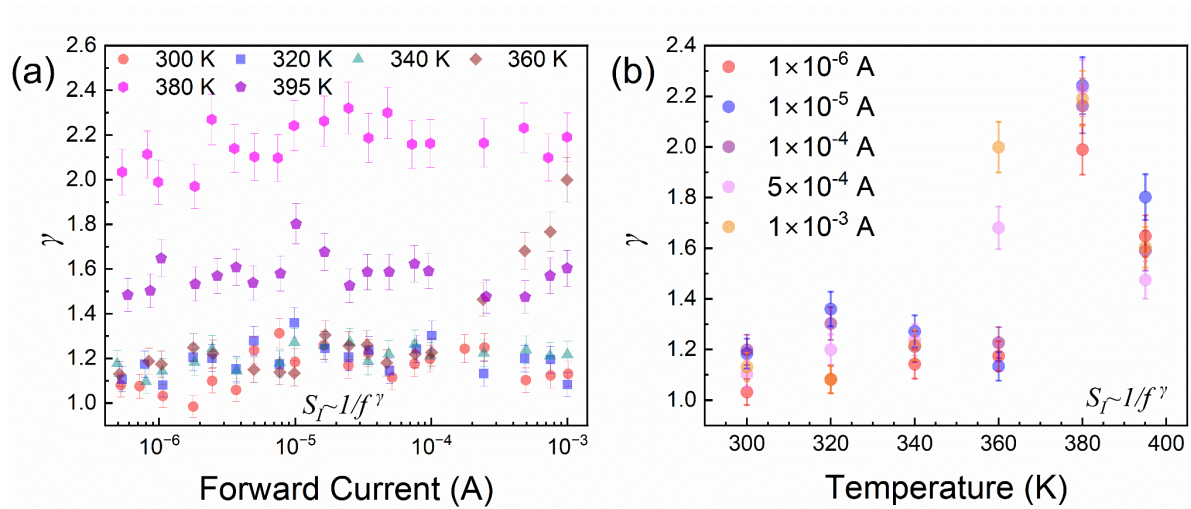

**Figure 3**: (a) The dependence of the noise parameter, $\gamma$, on the forward currents at different temperatures. The $\gamma$ value stays between 0.9 – 1.3 in the lower temperature region but moves towards a higher value of ~2 or more at 380 K due to the emergence of the Lorentzian component. (b) The $\gamma$ parameter dependence on temperature at different current levels.

The non-monotonic noise dependence on temperature is intriguing (see Figure 4 (b)). In the I–V characteristics, we did not observe any anomaly at 380 K (see Figure 1 (b)), which could explain the noise peak around that temperature. The noise increase at 380 K is accompanied by the appearance of the Lorentzian features, *i.e.*, $1/f^2$ tail. It is known that LFN can be extremely sensitive to various structure, morphological, defect density, and phase changes in the material. While I–V characteristics do not show any changes, the current fluctuations, *i.e.,* noise, can reveal even small structural or morphological variations. The LFN measurements were used to detect various phase transitions in different materials, including charge-density-wave phase transitions[39–41] and magnetic phase transitions in AFM materials[42]. The noise spectra at the transition temperatures generally show an increase in the noise level and an emergence of the pronounced Lorentzian features. Bulk single crystal $NiO_x$ reveals a magnetic phase transition from the AFM to the paramagnetic (PM) phase at the Néel temperature of 523 K[25]. However, the Néel temperature of $NiO_x$ can vary significantly depending on the size of NiO polycrystalline grains, morphology, stoichiometry, and lattice defects[43]. It was found that $NiO_x$ nanoparticles, with grain sizes of ~100 nm, have substantially lower Néel temperature, which can decrease even below 300 K as evidenced by the disappearance of the two-magnon peak in the Raman spectra[44].

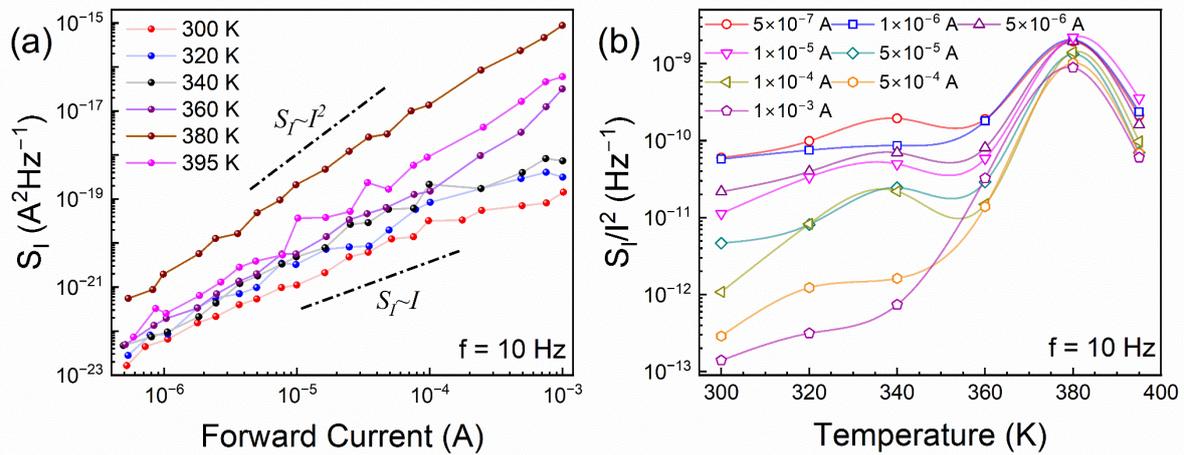

**Figure 4**: (a) The current noise spectral density, $S_I$, at $f$ = 10 Hz as a function of the forward current, $I$, at elevated temperatures. The noise initially follows $S_I \sim I$ dependence but changes to $S_I \sim I^2$ behavior at higher temperatures. (b) The normalized current noise spectral density, $S_I/I^2$, as a function of temperature, $T$, at different current values.

In order to detect any possible modifications in the material of the device upon heating, we conducted Raman spectroscopy measurements in the corresponding temperature range between 300 K and 398 K. The Raman measurements were conducted in the conventional backscattering configuration using λ = 488 nm laser excitation with 1800 g/mm grating and a 50× objective (Renishaw In-Via). The laser power was kept at ~3 mW to avoid local Joule heating in the device structure from the excitation laser. The diode was placed on top of a heating stage (Linkam THMS600) so that the Raman measurements can be conducted at elevated temperatures. Figure 5 (a) shows the temperature-dependent Raman spectra between 100 cm$^{-1}$ - 1800 cm$^{-1}$. The spectra are dominated by intense low-frequency phonon peaks from the β-Ga$_2$O$_3$ substrate. Only a small hump related to the two-magnon (2M) peak from the NiO$_x$ layer can be observed. The wave numbers of the observed spectral features between 100 cm$^{-1}$ and 800 cm$^{-1}$ are in good agreement with the literature on β-Ga$_2$O$_3$ Raman modes[45–47]. Figure 5 (b) shows the low-intensity peaks from the NiO$_x$ layer found between 700 cm$^{-1}$ - 1800 cm$^{-1}$. The spectra in that region contain the second-order LO and TO vibrational modes as well as the 2M peak related to the NiO material[25,48]. One can see that the 2M peak exists in the entire examined temperature range. This suggests that the NiO$_x$ layer remains in the AFM phase and the peak and Lorentzian spectral features observed in noise spectra are likely associated with other changes in materials used in NiO$_x$/β-Ga$_2$O$_3$ *p-n* heterojunction diodes.

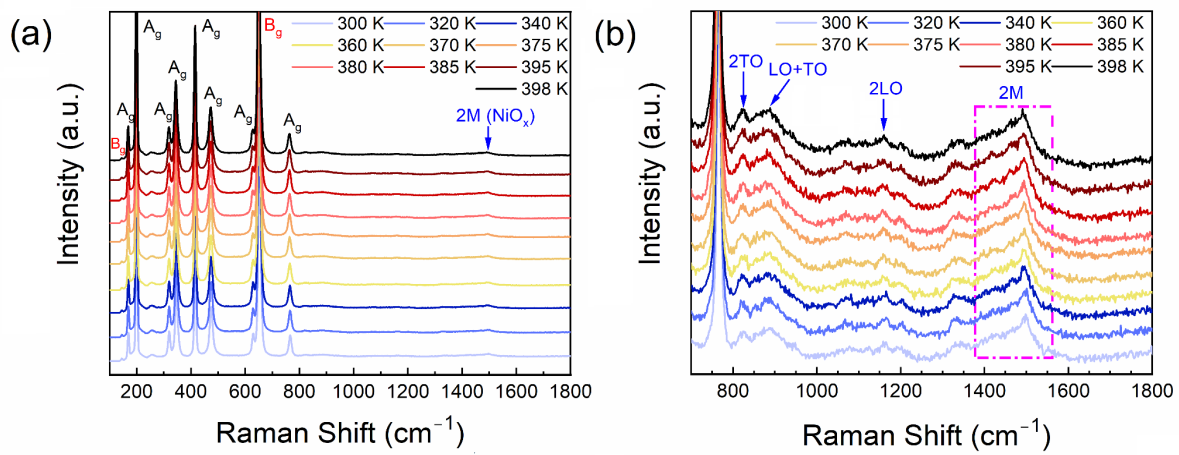

**Figure 5**: (a) Raman spectra of the NiO$_x$/β-Ga$_2$O$_3$ diode measured in the temperature range between 300 K to 398 K. The Raman spectra are dominated by the A$_g$ and B$_g$ vibrational modes of β-Ga$_2$O$_3$ substrate in the frequency range between 100 cm$^{-1}$ and 800 cm$^{-1}$. A small hump around 1500 cm$^{-1}$ is assigned to the two-magnon (2M) Raman peak of the antiferromagnetic NiO$_x$ layer. (b) The low-intensity peaks in the high-frequency region originate from the *p*-NiO$_x$ layer. One can see the second-order LO and TO phonon modes and the 2M peak, which remains visible in the entire measured temperature range.

To assess the possibility of the change in material characteristics, we plotted the Raman shifts and intensity as the functions of temperature for several peaks associated with the β-Ga$_2$O$_3$ substrates (see Figure 6 (a)-(d)). The Raman peak frequencies redshift with increasing temperature as expected for conventional semiconductor materials. The trend experiences a clear change at temperatures above 360 K (Figure 6 (a)-(b)). The temperature coefficients for different Raman peaks are different but the behavior around 360 K is consistent (see Supplementary Materials for additional data). We also observed a non-monotonic dependence of the peak intensities with a maximum value near 380 K (Figure 6 (c)-(d)). One can recall that the noise level also reached its maximum at ~ 380 K and then decreased (see Figure 4 (b)). The non-monotonic trend in the Raman intensity and the changes in the Raman temperature coefficients for the phonon modes associated with β-$Ga_2O_3$ and $NiO_x$/β-$Ga_2O_3$ heterointerface suggest a possibility of material changes at these temperatures, which result in reduced defects or improvement of the crystalline or morphological structure at $T > 380$ K.

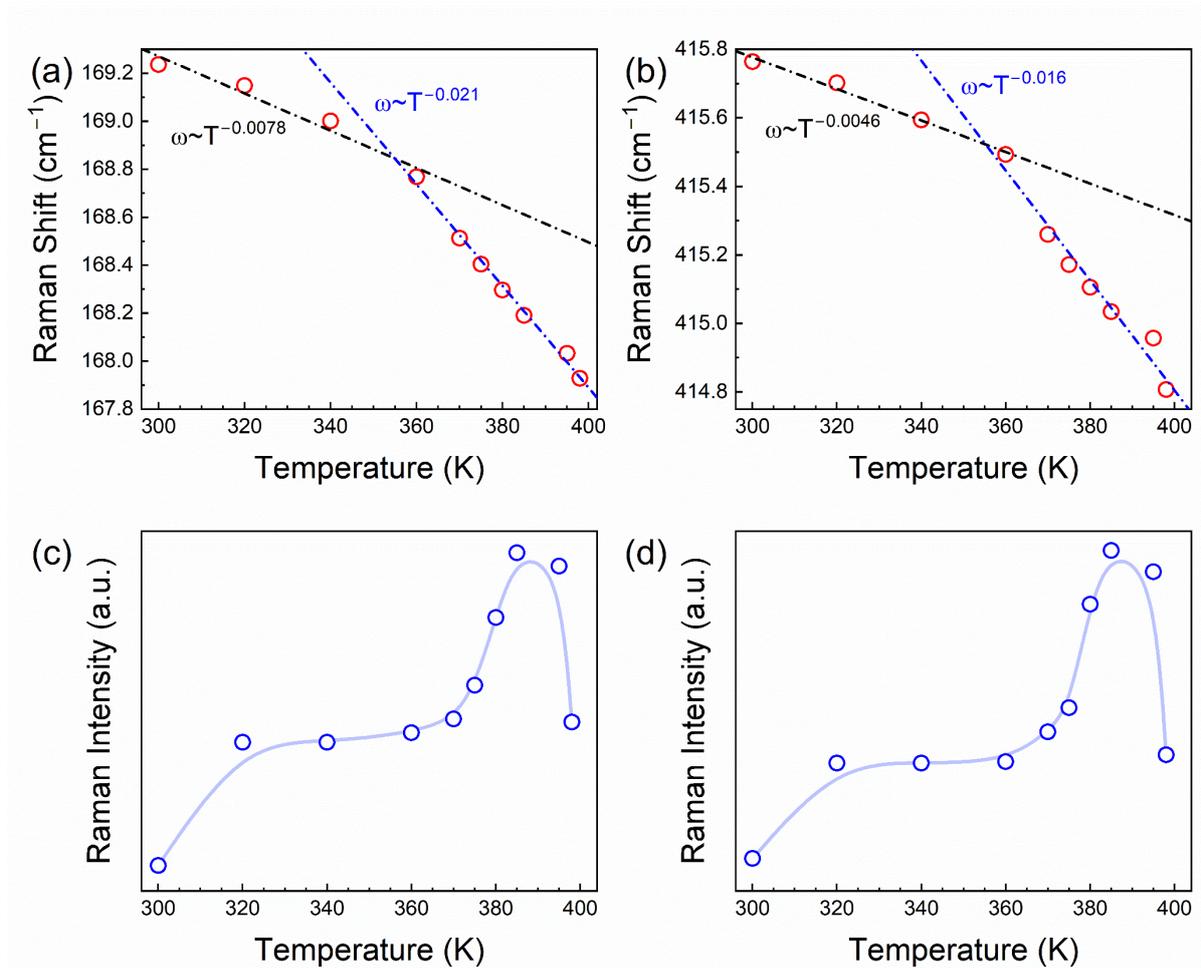

**Figure 6**: (a) Raman frequency as a function of temperature for the $A_g$ peak at ~169 cm$^{-1}$ from the β-$Ga_2O_3$ substrate. The frequency decreases with increasing temperature but changes more rapidly beyond 360 K. (b) The same as in (a) plotted for the $A_g$ Raman peak at ~415 cm$^{-1}$. Similar to $A_g$ peak shown in (a), the rate of change of the Raman frequency is larger after 360 K. (c) The Raman peak intensity as a function of temperature for the $A_g$ peak at ~169 cm$^{-1}$. The Raman intensity shows a non-monotonic dependence with the maximum value at $T = 385$ K. (d) The same as in (c) plotted for the $A_g$ Raman peak at ~415 cm$^{-1}$.

A recent study has shown that rapid thermal annealing of NiO$_x$/β-Ga$_2$O$_3$ heterojunction diodes can provide significant improvement in the quality of the *p-n* diode interface, accompanied by a decrease in the interfacial defect states acting as recombination centers[49]. In the reported study, annealing was performed at 225 °C in an N$_2$ atmosphere to improve Ga$_2$O$_3$/NiO interface quality[49]. The latter supports our suggestions that heating the device externally and increasing the temperature further due to the local Joule heating can change the material quality and result in the observed noise behavior. The emergence of the G-R Lorentzian bulges at $T \sim 380$ K can indicate the onset of changes in the material, followed by the noise decrease at higher temperatures because of the reduced number of defects contributing to *1/f* noise.

Another interesting feature in the noise response of NiO$_x$/β-Ga$_2$O$_3$ *p-n* heterojunction diodes is related to the noise spectral density dependence on current. It changes from linear, $S_I \sim I$, around 300 K to quadratic, $S_I \sim I^2$, near 380 K. It is not uncommon to observe distinct $S_I(I)$ behavior at different current regimes, as observed for other diode technologies[30–32,38,50,51]. Such noise behavior is attributed to different transport mechanisms dominating at different current regimes. It has been reported that the main transport mechanisms in NiO$_x$/β-Ga$_2$O$_3$ heterojunction diode are the interface recombination current or the trap-assisted tunneling current[52]. An interplay of these mechanisms may result in the changed noise spectral density dependence on current. A similar evolution of the $S_I(I)$ behavior was reported for other heterojunction devices[53,54]. For practical applications, it is important to compare the noise level in one electronic material and device type with that one in other electronic materials and devices. The noise level can be the overall indicator of the material quality and the maturity of the UWBG device technology. In Table I, we provide such a comparison for UWBG materials using our own and other reported data. One can see that the normalized noise spectral density of NiO$_x$/β-Ga$_2$O$_3$ *p-n* heterojunction diodes is on the order of $10^{-14}$ cm$^2$/Hz ($f = 10$ Hz) at 0.1 A/cm$^2$ current density, which is higher than that in GaN P-I-N diodes and GaN/AlGaN Schottky diodes, but lower than that in AlGaO Schottky diodes, diamond diodes, and comparable to the noise level in SiC *p-n* diodes.

**Table I:** Comparison of Noise Level in Wide-Bandgap Semiconductor Diode Technologies

| Device Type | $S_I$ (A$^2$Hz$^{-1}$) [$f$ = 10 Hz], $I$=10$^{-6}$ A | $S_I/I^2 \times Area$ (cm$^2$Hz$^{-1}$) [$f$ = 10 Hz] | Ref. |
|---|---|---|---|
| NiO$_x$/β-Ga$_2$O$_3$ $p$-$n$ diode | 6.5×10$^{-23}$ | 4×10$^{-14}$ | This work |
| GaN P-I-N diode | 10$^{-20}$ - 10$^{-22}$ | 10$^{-15}$ | [30] |
| Diamond diode | 10$^{-17}$ - 10$^{-18}$ | 10$^{-10}$ - 10$^{-12}$ | [31] |
| AlGaO Schottky diode | 1.5×10$^{-17}$ | 10$^{-12}$ | [32] |
| GaN/AlGaN Schottky diode | 10$^{-21}$ | 4×10$^{-15}$ | [38] |
| SiC $p$-$n$ diode | 10$^{-23}$ | 6×10$^{-14}$ | [50] |

## 4. Conclusion

In summary, we reported on the temperature dependence of LFN in NiO$_x$/β-Ga$_2$O$_3$ $p$-$n$ heterojunction diodes and compared the noise level in such devices with that in other UWBG technologies. The normalized noise spectral density in NiO$_x$/β-Ga$_2$O$_3$ $p$-$n$ heterojunction diodes is higher than that in GaN P-I-N diodes and GaN/AlGaN Schottky diodes, but lower than that in AlGaO Schottky diodes, diamond diodes, and comparable to the noise level in SiC $p$-$n$ diodes. We observed an intriguing non-monotonic dependence of the noise on temperature, which was attributed to the material characteristic changes. The Raman spectroscopy of the device structure suggests material changes, which results in reduced noise above this temperature. The obtained results are important for understanding the electronic characteristics of NiO$_x$/β-Ga$_2$O$_3$ heterojunctions and contribute to the development of noise spectroscopy as the quality assessment tool for new device technologies.

**Acknowledgments**

The work was supported by ULTRA, an Energy Frontier Research Center (EFRC) funded by the U.S. Department of Energy, Office of Science, Basic Energy Sciences under Award # DE-SC0021230.

**Conflict of Interest**

The authors declare no conflict of interest.

**Author Contributions**

A.A.B. coordinated the project and led the data analysis and manuscript preparation. S.G. conducted I–Vs, noise, and Raman measurements and contributed to the data analysis; D.H.M. fabricated *p-n* heterojunction diodes and measured I–Vs; Y.Z. and H.F. supervised device fabrication; F.K. contributed to the diode I–V and noise and Raman data analysis. All authors contributed to the manuscript preparation.

**The Data Availability Statement**

The data that support the findings of this study are available from the corresponding author upon reasonable request.